# Something new about radial wave functions of fermions in the repulsive Coulomb field

V.P.Neznamov [*] I.I.Safronov[†], V.E.Shemarulin[‡]

FSUE "RFNC-VNIIEF", Russia, Sarov, Mira pr., 37, 607188

Abstract

An impermeable barrier at $r = r_{cl}$ in the effective potential of the relativistic Schrödinger-type equation leads to exclusion of the range $0 \leq r < r_{cl}$ from the wave function domain. Based on duality of the Schrödinger-type equation and the Dirac equation, a similar exclusion should be made in the wave functions domain of the Dirac equation. As a result, we obtain new solutions to the Dirac equation in the Coulomb repulsive field.

Calculations show that depending on working parameters, at distances of several fractions or units of the Compton wavelength of the fermion from $r = r_{cl}$ new solutions almost coincide with the standard Coulomb functions of the continuous spectrum. Practically, matrix elements with new solutions will coincide to a good accuracy with standard matrix elements with the Coulomb functions of the continuous spectrum. Our consideration is methodological and helpful for discussing further development of quantum theory.

*Keywords: Dirac equation, second-order Schrödinger-type equation, Coulomb potential, effective potential, impermeable barrier, solutions of the Dirac equation and Schrödinger-type equation*

[*] vpneznamov@vniief.ru
vpneznamov@mail.ru
[†] IISafronov@vniief.ru
[‡] VEShemarulin@vniief.ru

# 1. Introduction

Quantum-mechanical motion of spin-½ particles in external electromagnetic and gravitational fields can be investigated in two ways, by using the Dirac equation with the bispinor wave function (see, for example, [1]) or the self-conjugate relativistic Schrödinger-type equation with the spinor wave function [2], [3]. Radial wave functions of both equations are related to each other by certain relations.

The necessity to use the relativistic Schrödinger-type equation arose when we were considering quantum mechanics of stationary states of particles in the spacetime of classical black holes.

In [4], the topic under study was interaction of scalar particles, photons, and fermions with the Schwarzschild, Reissner-Nordström, Kerr, and Kerr-Newman gravitational and electromagnetic fields with the zero and nonzero cosmological constant. Interaction of scalar particles, photons, and fermions with nonextremal rotating charged black holes in the minimal five-dimensional gauge supergravitation was also considered.

In almost all cases, the quantum-mechanical mode of a particle falling onto the event horizons is found to exist [5] - [7]. States of "a particle in field of classical black holes with zero-thickness event horizons" system are singular. The observed singularities do not allow full application of quantum theory, which makes it necessary to change the initial formulation of the physics problem.

An exception from the aforesaid for all considered black holes is the existence of degenerate stationary states of particles with energies $E = E^{st}$ [8] - [10].

When Dirac equations are used, solutions with $E = E^{st}$ are formally the only regular solutions, but they do not correspond to the real physical situation because of logarithmic divergence of Dirac normalization integrals in the vicinity of event horizons.

The situation changes when self-conjugate relativistic Schrödinger-type equations are used. As one passes from Dirac equations to Schrödinger-type equations, wave functions undergo nonunitary similarity transformations. As a result, wave functions become square-integrable [8] - [10]. Particles in states with $E = E^{st}$ are localized near event horizons.

Wave functions become zero at event horizons. Peaks of probability densities are fractions of the Compton wavelength of particles away from the event horizons depending on the working parameters of the systems under study.

In [11], we examined the results of using the Dirac equation and the Schrödinger-type equation in the flat Minkowski spacetime. The static Coulomb field was used as the external field.

In the case of the attractive Coulomb field, both approaches lead to coinciding energy spectra of spin-½ particles and similar probability densities [11].

However, in the case of the repulsive Coulomb field, the effective potential of the Schrödinger-type equation has a quantum-mechanical impermeable barrier at $r = r_{cl}$ [11]. In the region $r \geq r_{cl}$, solutions of the Schrödinger-type equation are Coulomb wave functions of the continuous spectrum [1] multiplied by a factor that ensures the self-conjugation of the equation. The region $0 \leq r < r_{cl}$ is inaccessible to particles whose motion is defined by the Schrödinger-type equation. Wave functions are zero at $r = r_{cl}$.

Despite the absence of the impermeable barrier, similar conclusions also follow for the wave functions of the Dirac equation in the repulsive Coulomb field since they are related to the wave functions of the Schrödinger-type equation. In this work, we investigate this seeming paradox in detail.

To analyze singularities and solutions of the radial Dirac equations, we use the Prüfer transformation and obtained equations for phase functions [12] - [15]. As a result, we arrived at new solutions of the Dirac equation with zero or finite values at $r = r_{cl}$. These solutions are related to the solutions of the Schrödinger-type equation. Interestingly, the new solutions come almost in coincidence with the analytical Coulomb functions of the continuous spectrum (29) at distances of a few fractions or units of the Compton wavelength of the fermion from $r = r_{cl}$.

The necessity of having correspondence between the solutions of the Dirac equation and the Schrödinger-type equation leads to exclusion of the region $0 \leq r < r_{cl}$ from physical consideration. When $r \to 0$, new solutions of the Dirac equation are divergent. Solutions of this type are considered physically unacceptable and are ignored in determination of the Coulomb functions of the discrete and continuous spectra. In our case, where the region $0 \leq r < r_{cl}$ is excluded from physical consideration, the problem of divergent solutions does not arise.

Below it is appropriate to mention the remarkable Foldy-Wouthuysen representation [16].

This representation is derived from the Dirac representation by a series of unitary transformations of the Dirac Hamiltonian and the Dirac wave function. Ultimately, we obtain two independent equations, one describing fermion states with positive energies and the other describing the ones with negative energies. An advantage of the Foldy-Wouthuysen representation is a transparent and unambiguous transformation from quantum mechanics operators to their classical equivalents [16] - [18].

Transition to the Foldy-Wouthuysen representation cannot be performed analytically. However, numerical calculations allow doing this with any required accuracy [19].

The Foldy-Wouthuysen representation is applied to consideration of quantum electrodynamics with calculation of some physical processes [20], [21]. The theory formalism is developed both for Dirac matrices in the Dirac-Pauli representation and for matrices in the spinor representation. Quantum electrodynamics effects are also considered using self-conjugate Schrödinger-type equation with spinor wave functions [22], [23].

In quantum electrodynamics versions with spinors in fermion equations, there is no need for the vacuum polarization concept, and, consequently, there are no processes of vacuum creation and annihilation of particle - antiparticle pairs [24].

In this work, we focus on the analysis of solutions to the Dirac equations with the bispinor wave function and to the self-conjugate relativistic Schrödinger-type equation with the spinor wave function in the external repulsive Coulomb field.

Here the electron with the charge $e < 0$ is viewed as a fermion.

The work is organized as follows. Section 2 presents equations for describing quantum-mechanical motion of electrons in the external repulsive Coulomb field. For the Schrödinger-type equation, occurrence, and properties of the impermeable barrier at $r = r_{cl}$ are discussed. Section 3 considers we discuss the asymptotics of the radial functions of the Dirac equation and the Schrödinger-type equation in the vicinity of the impermeable barrier. Section 4 deals with analysis of singularities and numerical solutions of the Dirac equation. The Conclusions Section summarizes the main results.

## 2. Equations for description of electron motion in the external Coulomb field

### *2.1 The Dirac equation*

The system of units $\hbar = c = 1$ is mainly used below. The Dirac equation has the form

$$i\frac{\partial \psi}{\partial t} = \left(\boldsymbol{\alpha}\mathbf{p} + \beta m + V(r)\right)\psi, \qquad (1)$$

where $\mathbf{p} = -i\dfrac{\partial}{\partial \mathbf{r}}$, $\alpha^k, \beta$ are the four-dimensional Dirac matrices, $m$ is the electron mass, $V(r) = \dfrac{Ze^2}{r}$ is the repulsive Coulomb potential, $Z$ is formally the ordinal number of the atomic antinucleus with the charge $-Z|e|$, $\psi(\mathbf{r},t) = \begin{pmatrix}\varphi(\mathbf{r},t)\\\chi(\mathbf{r},t)\end{pmatrix}$ is the bispinor, $\varphi(\mathbf{r},t), \chi(\mathbf{r},t)$ are the spinor wave functions.

In case of centrally symmetric external fields, one can separate angular and radial variables. In spherical coordinates for stationary states, we write the bispinor $\psi$ as [25]

$$\psi(r,\theta,\varphi,t) = \begin{pmatrix} F(r)\xi(\theta) \\ -iG(r)\sigma^3\xi(\theta) \end{pmatrix} e^{-iEt} e^{im_\varphi \varphi}, \quad (2)$$

where $E$ is the electron energy, $\xi(\theta)$ are the spherical harmonics for spin-½ particles, $m_\varphi = -j, -j+1, ..., j$; $j$ is the quantum number of the total particle momentum, and $\sigma^3$ is the two-dimensional Pauli matrix.

To separate the variables, it is convenient to make the equivalent replacement $\alpha^1 \to \alpha^3, \alpha^2 \to \alpha^1, \alpha^3 \to \alpha^2$. In addition, we introduce dimensionless variables $\varepsilon = E/m$, $\rho = r/l_c$, $l_c = 1/m$ is the Compton wavelength of the electron, and $\alpha_{fs} = e^2$ is the electromagnetic fine-structure constant. Then the system of the Dirac equations for the radial wave functions and the repulsive Coulomb potential $V(\rho) = Z\alpha_{fs}/\rho$ has the form (see, for example, [1])

$$\left(\varepsilon + 1 - \frac{Z\alpha_{fs}}{\rho}\right)G(\rho) = \frac{dF}{d\rho} + \frac{1+\kappa}{\rho}F(\rho), \quad (3)$$

$$-\left(\varepsilon - 1 - \frac{Z\alpha_{fs}}{\rho}\right)F(\rho) = \frac{dG}{d\rho} + \frac{1-\kappa}{\rho}G(\rho). \quad (4)$$

In (3) and (4), $\kappa$ is the quantum number of the Dirac equation,

$$\kappa = \mp 1, \mp 2... = \begin{cases} -(l+1), & j = l+1/2, \\ l, & j = l-1/2, \end{cases}$$

where $j, l$ are the quantum numbers of the total and orbital momenta of the spin-½ particle.

If we introduce the phase function [12] - [15]

$$\tan\Phi(\rho) = \frac{F(\rho)}{G(\rho)}, \quad (5)$$

$$P^2(\rho) = F^2(\rho) + G^2(\rho). \quad (6)$$

$$F(\rho) = P(\rho)\sin\Phi(\rho), \quad (7)$$

$$G(\rho) = P(\rho)\cos\Phi(\rho), \quad (8)$$

we can obtain from (3) and (4) equations for $\Phi(\rho)$ and $P(\rho)$

$$\frac{d\Phi}{d\rho} = \varepsilon - \frac{Z\alpha_{fs}}{\rho} + \cos(2\Phi) - \frac{\kappa}{\rho}\sin(2\Phi), \quad (9)$$

$$\frac{d\ln P}{d\rho} = -\frac{1}{\rho} + \frac{\kappa}{\rho}\cos(2\Phi) + \sin(2\Phi). \tag{10}$$

The radial probability density for the Dirac equation is

$$w_D = \left(F^2(\rho) + G^2(\rho)\right)\rho^2. \tag{11}$$

## 2.2 The second-order Schrödinger-type equation

First, one should square the Dirac equations, then go from the bispinor to the spinor wave functions, and reduce the equations to the self-conjugate form [3], [11].

As is known, self-conjugate second-order equations have some remarkable properties that distinguish them among equations of the general form. In particular, eigenfunctions of boundary-value problems for self-conjugate equations make up complete orthonormalized systems in the space of doubly continuously differentiable functions, i.e., any function from this space is expandable in a regularly convergent Fourier series in eigenfunctions (see, for example, [26], [27]).

For the centrally symmetric Coulomb potential, the thus obtained equations allow separation of variables in the spherical coordinates $(r, \theta, \varphi)$. The Schrödinger-type equations (of second order) with effective potentials $U_{\mathit{eff}}^F(r), U_{\mathit{eff}}^G(r)$ have the form [11]

$$\frac{d^2 F_{Schr}(\rho)}{d\rho^2} + 2\left(E_{Schr} - U_{\mathit{eff}}^F(\rho)\right)F_{Schr}(\rho) = 0, \tag{12}$$

$$\frac{d^2 G_{Schr}(\rho)}{d\rho^2} + 2\left(E_{Schr} - U_{\mathit{eff}}^G(\rho)\right)G_{Schr}(\rho) = 0, \tag{13}$$

where

$$E_{Schr} = \left(\varepsilon^2 - 1\right)/2. \tag{14}$$

In (12), (13),

$$U_{\mathit{eff}}^F(\rho) = \varepsilon V - \frac{1}{2}V^2 + \frac{\kappa(\kappa+1)}{2\rho^2} + \frac{1}{4}\frac{d^2V/d\rho^2}{\varepsilon+1-V} + \frac{3}{8}\frac{(dV/d\rho)^2}{(\varepsilon+1-V)^2} - \frac{1}{2}\frac{\kappa\, dV/d\rho}{\rho(\varepsilon+1-V)}. \tag{15}$$

$$U_{\mathit{eff}}^G(\rho) = \varepsilon V - \frac{1}{2}V^2 + \frac{\kappa(\kappa+1)}{2\rho^2} + \frac{1}{4}\frac{d^2V/d\rho^2}{\varepsilon-1-V} + \frac{3}{8}\frac{(dV/d\rho)^2}{(\varepsilon-1-V)^2} - \frac{1}{2}\frac{\kappa\, dV/d\rho}{\rho(\varepsilon-1-V)}. \tag{16}$$

For self-conjugate equations (12) and (13), radial probability densities are

$$w_F = F_{Schr}^2(\rho), \tag{17}$$

$$w_G = G_{Schr}^2(\rho). \tag{18}$$

Equations (12) and (13) can be used for describing motion of particles and antiparticles respectively. Below, to describe electron motion, we will use Eq. (12) with the effective potential $U_{eff}^{F}(\rho)$ (15).

Unlike the Coulomb potential $V = Z\alpha_{fs}/\rho$, the effective potential $U_{eff}^{F}(\rho)$ (15) is singular (see Fig.1) for

$$\rho_{cl} = Z\alpha_{fs}/(\varepsilon+1) \tag{19}$$

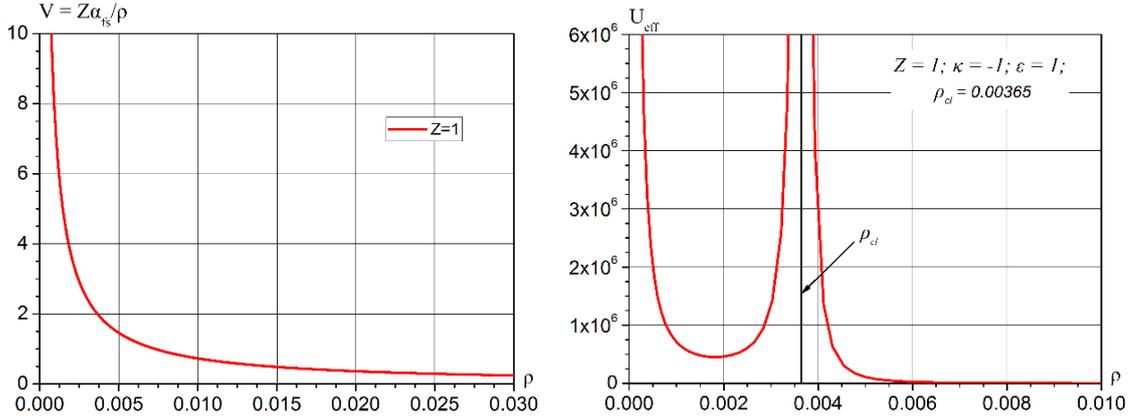

Fig. 1. Dependences $V(\rho)$ and $U_{eff}(\rho)$.

In natural units, $r_{cl}$ is proportional to the classical electron radius $r_f = Ze^2/mc^2$

$$r_{cl} = \frac{r_f}{1+E/mc^2} = \frac{Z(e^2/mc^2)}{1+E/mc^2}. \tag{20}$$

The leading asymptotics of potential (15) when $\rho \to \rho_{cl}$ is

$$U_{eff}^{F}\bigg|_{\rho \to \rho_{cl}} = \frac{3}{8}\frac{1}{(\rho-\rho_{cl})^2}. \tag{21}$$

Potential barrier (21) is quantum-mechanically impermeable [28]. The quantum particle cannot penetrate the region $\rho < \rho_{cl}$. Exclusion of the region $0 \leq \rho < \rho_{cl}$ from physical consideration does not effect the calculated cross section of electron scattering in the Coulomb field [11] and the interpretation of the experimental results on probing the internal structure of the electron [29], [11].

## 3. Asymptotics of functions in the region of the impermeable barrier

### *3.1 The Schrödinger-type equation*

If the radial function $F_{Schr}(\rho)$ with $\rho \to \rho_{cl}$ is written as

$$F_{Schr}(\rho)\big|_{\rho \to \rho_{cl}} = (\rho - \rho_{cl})^s \sum_{k=0}^{\infty} f_k (\rho - \rho_{cl})^k, \tag{22}$$

the indicial equation for (12) with allowance for (21) leads to two solutions of $s_1 = 3/2, s_2 = -1/2$, and

$$(F_{Schr})_1\big|_{\rho \to \rho_{cl}} = f_0^{(1)} (\rho - \rho_{cl})^{3/2}, \tag{23}$$

$$(F_{Schr})_2\big|_{\rho \to \rho_{cl}} = f_0^{(2)} (\rho - \rho_{cl})^{-1/2}. \tag{24}$$

Solution (24) diverges when $\rho \to \rho_{cl}$ and is non-normalizable. Similar solutions are physically unacceptable. Similarly, in the case of solving the Dirac equation in the attractive and repulsive Coulomb fields, physically unacceptable solutions are $F(\rho), G(\rho) \sim \dfrac{1}{\rho^{\gamma}}$ diverging when $\rho \to 0$ (see, for example, [1]). Here $\gamma = \sqrt{\kappa^2 - Z^2 \alpha_{fs}^2}$.

### *3.2 Dirac equation*

The radial wave functions $F_{Schr}(\rho)$, $G_{Schr}(\rho)$ of the Schrödinger-type equation are connected with the wave functions $F(\rho), G(\rho)$ of the Dirac equation by the relations (see [11])

$$F_{Schr}(\rho) = \left(\varepsilon + 1 - \frac{Z\alpha_{fs}}{\rho}\right)^{-1/2} \rho F(\rho), \tag{25}$$

$$G_{Schr}(\rho) = \left(|\varepsilon| + 1 + \frac{Z\alpha_{fs}}{\rho}\right)^{-1/2} \rho G(\rho). \tag{26}$$

On the basis of (23) and (24), and in accordance with (3) and (4), in the vicinity of $\rho_{cl}$ there should exist two solutions of the Dirac equation

$$\begin{cases} F_1\big|_{\rho \to \rho_{cl}} \sim (\rho - \rho_{cl})^2, \\ G_1\big|_{\rho \to \rho_{cl}} \sim \text{const 3}, \end{cases} \tag{27}$$

$$\begin{cases} F_2\big|_{\rho \to \rho_{cl}} \sim \text{const 1}, \\ G_2\big|_{\rho \to \rho_{cl}} \sim \text{const 2}. \end{cases} \tag{28}$$

To solutions (28) there correspond the Coulomb functions of the continuous spectrum with $\varepsilon > 1$.

Wave functions in the repulsive Coulomb field normalized to divergent spherical waves when $\rho \to \infty$ can be written in dimensionless variables as [1]

$$\left.\begin{matrix} F(\rho) \\ G(\rho) \end{matrix}\right\} = 2^{3/2} \sqrt{\frac{\varepsilon \pm 1}{\varepsilon}} e^{\frac{\pi v}{2}} \frac{|\Gamma(\gamma + 1 + iv)|}{\Gamma(2\gamma + 1)} \frac{(2p\rho)^{\gamma}}{\rho} \times \\ \times \begin{matrix} \mathrm{Im} \\ \mathrm{Re} \end{matrix} \left\{ e^{i(p\rho + \xi)} F(\gamma - iv, 2\gamma + 1, -2ip\rho) \right\}, \qquad (29)$$

where $F(\alpha, \beta, z)$ is the degenerated hypergeometric function, $\Gamma(z)$ is the gamma function,

$\gamma = \sqrt{\kappa^2 - Z^2 \alpha_{fs}^2}$, $p = \sqrt{\varepsilon^2 - 1}$, $v = -\dfrac{Z\alpha_{fs}\varepsilon}{p}$, and $e^{-2i\xi} = \dfrac{\gamma - iv}{\kappa - iv/\varepsilon}$.

Functions (29) at $\rho = \rho_{cl}$ are equal to constants. This behavior leads to nonphysical asymptotics (24) for the wave function of the Schrödinger-type equation.

Asymptotics (27) do not conflict with the system of Dirac equations (3) and (4). However, solutions with asymptotics (27) have not been reported in the literature so far. We will obtain these solutions using (5) - (10).

## 4. Numerical solutions of the Dirac equations for the phase function

### *4.1 The singular points of equation (9)*

At the singular points $(\rho, \Phi)$ of Eq. (9)

$$\frac{d\Phi}{d\rho} = \frac{\rho(\varepsilon + \cos 2\Phi) - Z\alpha_{fs} - \kappa \sin 2\Phi}{\rho}$$

the numerator and the denominator should become zero

$$\rho = 0, \quad \kappa \sin 2\Phi = -Z\alpha_{fs}. \qquad (30)$$

Consequently, these singular points exist only if

$$\left| -\frac{Z\alpha_{fs}}{\kappa} \right| < 1. \qquad (31)$$

Dirac's theory with the point Coulomb potential is valid only for $Z \le 137$. In this case, the equality $\left| -Z\alpha_{fs}/\kappa \right| = 1$ is not achieved.

For $Z > 137$, due to excessive singularity of the Coulomb potential when $\rho \to 0$, a quantum-mechanical problem of falling onto the center arises. To eliminate the problem, self-conjugate extension of the Hamiltonian is needed, which is eventually reduced to considering the finite size of atomic nuclei (see, for example, [11]).

From the second equality in (30) it follows that

$$\Phi_k = \frac{(-1)^{k+1}}{2}\arcsin\frac{Z\alpha_{fs}}{\kappa} + \frac{\pi}{2}k, \quad k = 0,\pm1,\pm2,... \qquad (32)$$

Solutions of (9) are periodic with period $\pi$, and singular points $(0,\Phi_k)$ are "quasi-periodic" with period $\pi/2$ $\left(\Phi_k + \frac{\pi}{2} \approx \Phi_{k+1}\right)$.

Let $Z = 1, \kappa = -1$. Then in the period $\left[-\frac{\pi}{2}, \frac{\pi}{2}\right]$

$$\Phi_0 \approx 0.00365, \quad \Phi_1 \approx \frac{\pi}{2} - 0.00365. \qquad (33)$$

For $Z = 1, \kappa = 1$, and in the period $\left[-\frac{\pi}{2}, \frac{\pi}{2}\right]$

$$\Phi_0 \approx -0.00365, \quad \Phi_{-1} \approx -\frac{\pi}{2} + 0.00365. \qquad (34)$$

Similar forms of $\Phi_0, \Phi_1$ and $\Phi_0, \Phi_{-1}$ can be given for any negative and positive values of $\kappa$ and for arbitrary values of $Z \geq 1$.

Note that when $Z$ increases, the singular points $(0,\Phi_0),(0,\Phi_1)$ and $(0,\Phi_0),(0,\Phi_{-1})$ in the period $\left[-\frac{\pi}{2}, \frac{\pi}{2}\right]$ approach one another but do not coincide because inequality (31) holds.

### *4.2 On the character of the singular points of equation (9)*

Let us expand the numerator and the denominator of the right-hand side of (9) in the vicinity of the singular point $(0,\Phi_k)$ up to the terms of second order in $\rho$ and $(\Phi - \Phi_k)$

$$\frac{d\Phi}{d\rho} = \frac{\rho(\varepsilon + \cos 2\Phi) - Z\alpha_{fs} - \kappa\sin 2\Phi}{\rho} = \frac{c\rho + d(\Phi - \Phi_k) + P_1(\rho,\Phi)}{a\rho + b(\Phi - \Phi_k)},$$

$a = 1, \quad b = 0, \quad c = \varepsilon + \cos(2\Phi_k), \quad d = -2\kappa\cos(2\Phi_k),$

$P_1(\rho,\Phi) = -2\sin(2\Phi_k)\rho(\Phi - \Phi_k) + 2\kappa\sin(2\Phi_k)(\Phi - \Phi_k)^2 + ...$

The character of the singular points is determined by the roots $\lambda_1, \lambda_2$ of the characteristic Eq. ([30], pp. 441-442; [31], Chapter 6)

$$\begin{vmatrix} a-\lambda & b \\ c & d-\lambda \end{vmatrix} = 0, \quad \lambda_1 = 1, \quad \lambda_2 = -2\kappa\cos(2\Phi_k).$$

Now we consider the determinant $\Delta_k = \begin{vmatrix} a & b \\ c & d \end{vmatrix} = \lambda_1\lambda_2 = -2\kappa\cos(2\Phi_k).$

To classify the singular points, we first clarify whether the equalities $\lambda_1 = \lambda_2$ and $\Delta_k = 0$ are possible. It is sufficient to consider singular points in the period $\left[-\dfrac{\pi}{2}, \dfrac{\pi}{2}\right]$, where $\cos(\Phi) \geq 0$. In this case,

$$\lambda_2 = -2\kappa \cos(2\Phi_k) = -2\kappa \cos\left((-1)^{k+1} \arcsin\dfrac{Z\alpha_{fs}}{\kappa} + k\pi\right) = 2\,\mathrm{sign}(\kappa)(-1)^{k+1}\sqrt{\kappa^2 - (Z\alpha_{fs})^2}.$$

If $\kappa < 0$ and $k$ is odd or $\kappa > 0$ and $k$ is even, then $\lambda_2 < 0$ and, consequently, $\lambda_2 \neq \lambda_1$.

If $\kappa < 0$ and $k$ is even or $\kappa > 0$ and $k$ is odd, then $\lambda_2 = 2\sqrt{\kappa^2 - (Z\alpha_{fs})^2} > 0$. The difference $\kappa^2 - (Z\alpha_{fs})^2 > 0$ is not an integer. Consequently, the expression $\sqrt{\kappa^2 - (Z\alpha_{fs})^2}$ is irrational; and $\lambda_2 \neq \lambda_1$. In addition, $\Delta_k = 2\,\mathrm{sign}(\kappa)(-1)^{k+1}\sqrt{\kappa^2 - (Z\alpha_{fs})^2} \neq 0$.

Thus, for the singular points $(0, \Phi_k)$ of Eq. (9), the inequalities $\lambda_1 \neq \lambda_2$ and $\Delta_k \neq 0$ always hold, and the character of these points is determined by the sign of the determinant $\Delta_k = -2\kappa \cos(2\Phi_k) = 2\,\mathrm{sign}(\kappa)(-1)^{k+1}\sqrt{\kappa^2 - (Z\alpha_{fs})^2}$, or, equivalently, by the sign of the product $\mathrm{sign}(\kappa)(-1)^{k+1}$ ([16], [17]).

1. If $\Delta_k = -2\kappa \cos(2\Phi_k) > 0$, the point $(0, \Phi_k)$ is a node.

2. If $\Delta_k = -2\kappa \cos(2\Phi_k) < 0$, the a point $(0, \Phi_k)$ is a saddle.

Hence it follows that a change in the sign of $\kappa$ entails a change in the character of the point $(0, \Phi_k)$: a node becomes a saddle, and a saddle becomes a node. This is illustrated by Figs. 2 and 3.

### *4.3 Integral curves of equations* (9)

Integral curves of the equation for the phase function were analyzed both by numerically solving Eq. (9) and by using the Maple code with the expansion of solutions of (9) up to $\rho^3$. The results obtained in both cases agree with the results of the theoretical analysis in Section 4.2.

For negative $\kappa$, the singular point $(0, \Phi_1)$ is a saddle ($\mathrm{sign}(\kappa)(-1)^{1+1} < 0$)) with a separatrix analytically coinciding with the Coulomb function $F(\rho)$, and the singular point $(0, \Phi_0)$ is a node ($\mathrm{sign}(\kappa)(-1)^{0+1} > 0$). For $\kappa > 1$, the singular point $(0, \Phi_0)$ is a saddle $(\text{sign}(\kappa)(-1)^{0+1} < 0)$, and the point $(0, \Phi_{-1})$ is a node $(\text{sign}(\kappa)(-1)^{-1+1} > 0)$. The cases for $\kappa = \mp 1$, $Z = 1$, $\varepsilon = 1.2$ are shown in Figs. 2 and 3.

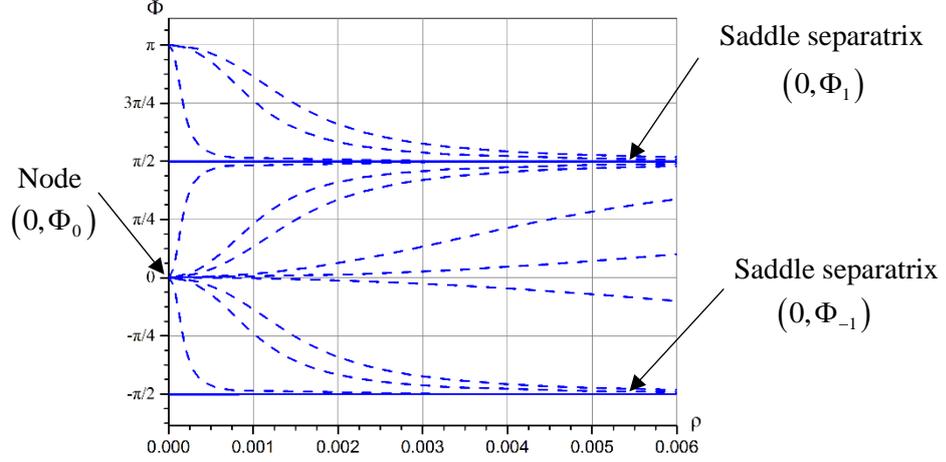

Fig. 2. Portrait of integral curves in the region of singular points for $\kappa = -1$: $\Phi_0 \approx 0.00365$, $\Phi_1 \approx \pi/2 - 0.00365$ and $\Phi_{-1} \approx -\pi/2 - 0.00365$.

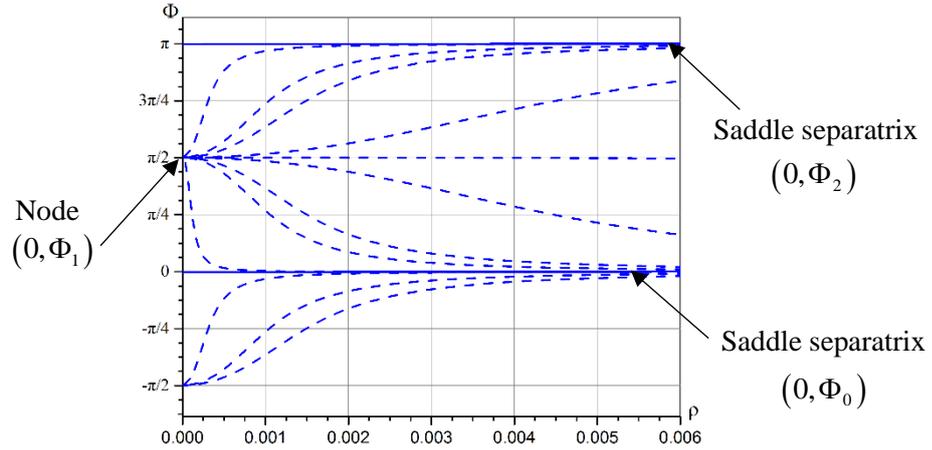

Fig. 3. Portrait of integral curves in the region of singular points for $\kappa = +1$: $\Phi_0 \approx -0.00365$, $\Phi_{-1} \approx -\pi/2 + 0.00365$ and $\Phi_2 \approx \pi - 0.00365$.

A general solution of (9) when $\rho \to 0$ has the form

$$\Phi = \text{arctg}\left(\frac{1}{Z\alpha_{fs}}\left(\frac{2\gamma}{1 - C\rho^{-2\gamma}} - \kappa - \gamma\right)\right) + k\pi, \quad k = 0, \pm 1, \pm 2. \tag{35}$$

The integration constant $C$ parameterizes family of integral curves depicted in Fig. 2 and 3. Irrespective of the sign of the parameter $\kappa$, the constant corresponding to the node separatrix is $C = \pm\infty$, and the constant corresponding to the saddle separatrix is $C = 0$; the curves with negative $C$ go up from the node (the larger the constant, the higher the curve), and the curves with positive $C$ go down (as in the case of negative $C$, the larger the constant, the higher the curves).

The curves starting at the node may be upper or lower adjacent to the saddle separatrices. There are also curves starting at the node that are not close to the saddle separatrices up to $\rho \approx 10$.

The Dirac functions $F(\rho) = P(\rho)\sin\Phi(\rho), G(\rho) = P(\rho)\cos\Phi(\rho)$ corresponding to the curves $\Phi(\rho)$ going into the node diverge as $1/\rho^\gamma$ when $\rho \to 0$, where $\gamma = \sqrt{\kappa^2 - \alpha_{fs}^2 Z^2}$. This family of solutions is usually rejected as nonphysical when determining the discrete spectrum of hydrogen-like atoms and when obtaining wave functions of the discrete and continuous spectra.

In our case, based upon duality of solutions to the Schrödinger-type equation and the Dirac equations, the region $0 \leq \rho < \rho_{cl}$ should necessarily be excluded from consideration for both equations (see Sec. 2.2 and (25), (26) in Sec. 3.2).

For the region $\rho \geq \rho_{cl}$, there is no $\sim 1/\rho^\gamma$ divergence problem for the Dirac functions $F(\rho), G(\rho)$. Among the phase curves going into the node there is always our necessary curve with $\Phi(\rho_{cl}) = 0$ and $\Phi \sim (\rho - \rho_{cl})^2$ where $\rho \to \rho_{cl}$. It is numerically found by integrating Eq. (9), from $\rho = \rho_{cl}$ to $\rho \approx 0$.

### *4.4 Numerical solutions*

Coulomb functions (29) with $\varepsilon > 1$ were also obtained as a solution of Dirac equations (9) and (10).

Figure 4 shows functions (29) expressed in terms of the dependence $\tan\Phi(\rho) = F(\rho)/G(\rho)$. This dependence is obtained using the Maple code. The figure also shows a similar dependence obtained by numerical left-to-right integration of Eq. (9) with the boundary condition from formulas (29) $\tan\Phi(0) = F(0)/G(0)$. The numerical and analytical solutions are seen to be in good agreement.

Figure 5 shows the numerical solution of (9) with the boundary condition $\Phi(\rho_{cl}) = 0$ for the same parameters as in Fig. 4. For comparison, the analytical solution of (29) in the form $\Phi(\rho) = \arctan\dfrac{F(\rho)}{G(\rho)}$ is also shown.

The difference between the solutions confirms that for the repulsive Coulomb potential there exists the second solution of the Dirac equation with asymptotics (27). It is remarkable that at the distance less than one Compton wavelength from $\rho = \rho_{cl}$ this solution is very close to the analytical function $F(\rho)$ of the continuous spectrum (29).

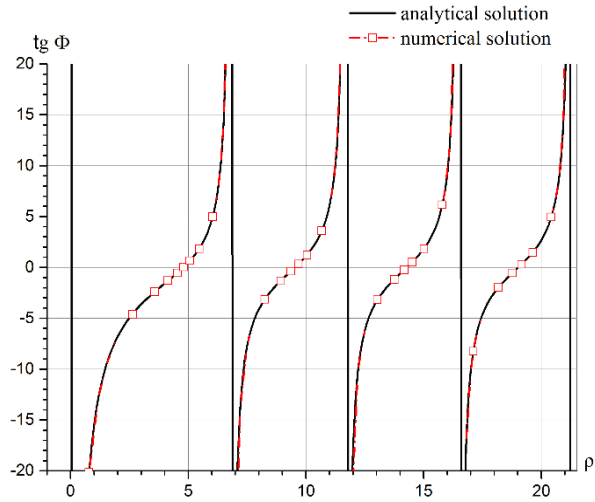

Fig. 4. Dependences $\tan\Phi(\rho) = F(\rho)/G(\rho)$ for the parameters of $Z = 1$, $\varepsilon = 1.2$, $\kappa = -1$.

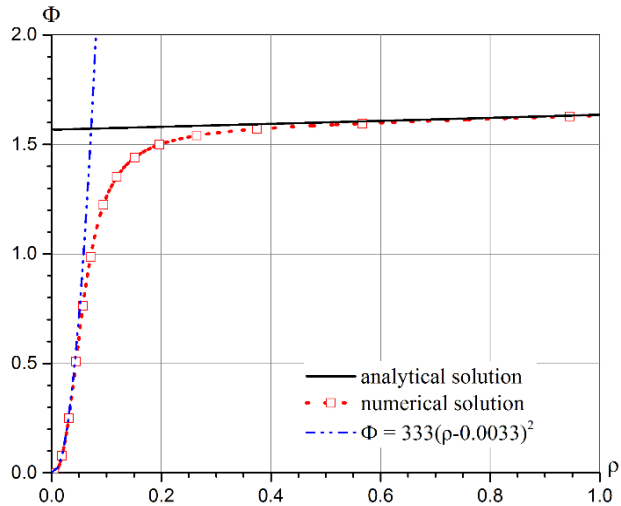

Fig. 5. Analytical and numerical solutions for $\Phi(\rho)$ near $\rho_{cl} = 0.0033$.

Figures 6 – 11 present calculated dependences $\Phi(\rho), F(\rho), G(\rho)$ for the calculation parameters $\varepsilon = 1.2$; $\kappa = -1, +1$; $Z = 1, 10, 137$. All of them are obtained with the boundary condition $\Phi(\rho_{cl}) = 0$. For comparison, these corresponding dependences obtained from (29) are also given in these figures.

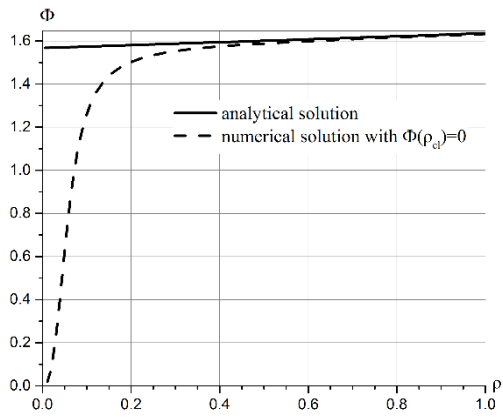
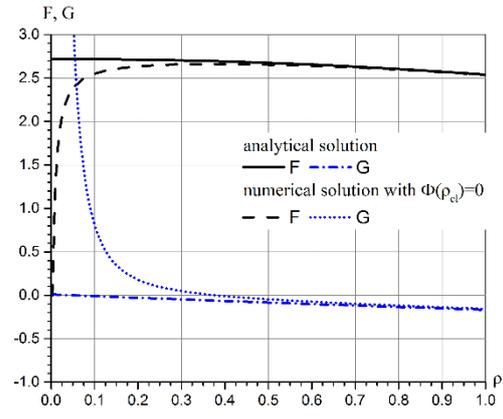

a) function $\Phi(\rho)$

b) function $F(\rho), G(\rho)$ ($G(\rho_{cl}) = 752$ for the case of $\Phi(\rho_{cl}) = 0$)

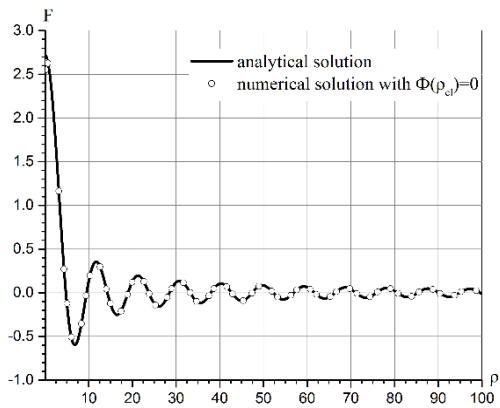
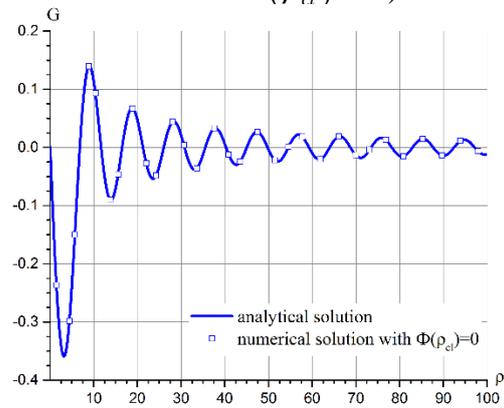

c) function $F(\rho)$

d) function $G(\rho)$

Fig. 6. Dependences $\Phi(\rho), F(\rho), G(\rho)$ for the parameters
$\varepsilon = 1.2, \ \kappa = -1, \ Z = 1, \ \rho_{cl} = 0.003317$.

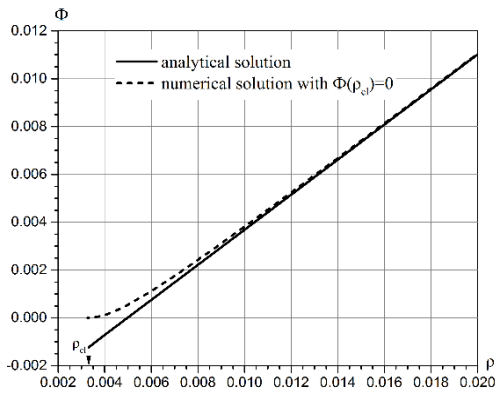
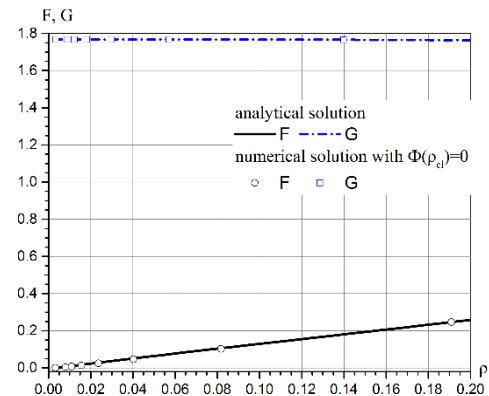

a) function $\Phi(\rho)$

б) function $F(\rho), G(\rho)$

Fig. 7. Dependences $\Phi(\rho), F(\rho), G(\rho)$ for the parameters
$\varepsilon = 1.2, \ \kappa = +1, \ Z = 1, \ \rho_{cl} = 0.003317$.

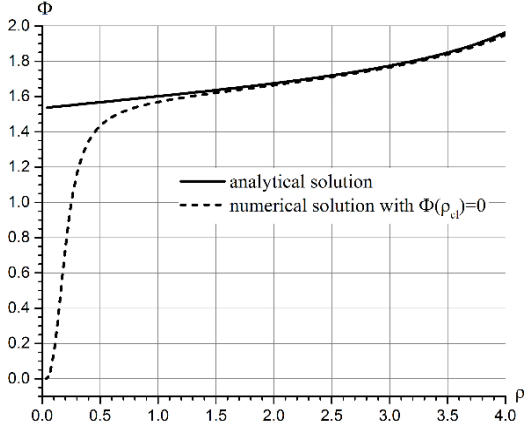
a) function $\Phi(\rho)$

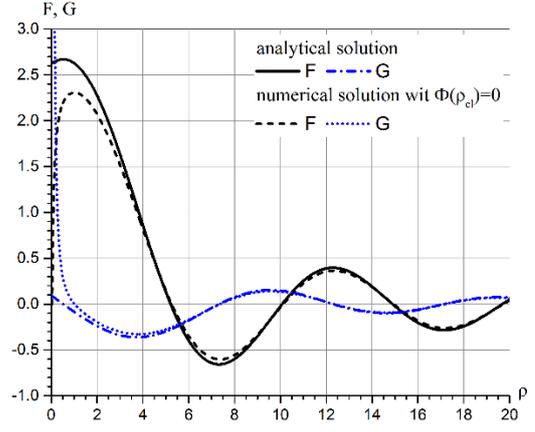
b) function $F(\rho), G(\rho)$ ($G(\rho_{cl})=68$ for the case of $\Phi(\rho_{cl})=0$)

Fig. 8. Dependences $\Phi(\rho), F(\rho), G(\rho)$ for the parameters
$\varepsilon = 1.2$, $\kappa = -1$, $Z = 10$, $\rho_{cl} = 0.03317$.

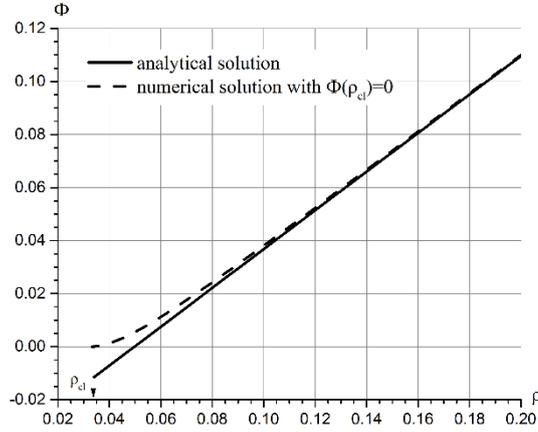
a) function $\Phi(\rho)$

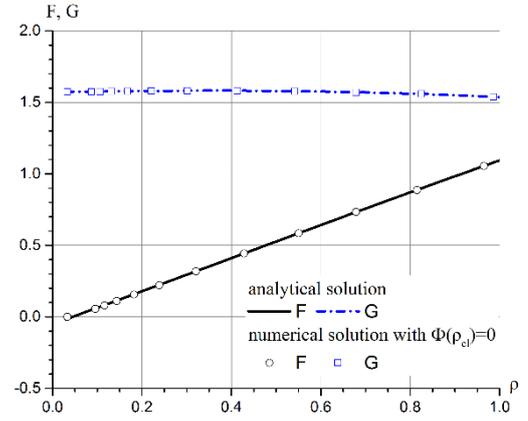
b) functions $F(\rho), G(\rho)$

Fig. 9. Dependences $\Phi(\rho), F(\rho), G(\rho)$ for the parameters
$\varepsilon = 1.2$, $\kappa = +1$, $Z = 10$, $\rho_{cl} = 0.03317$.

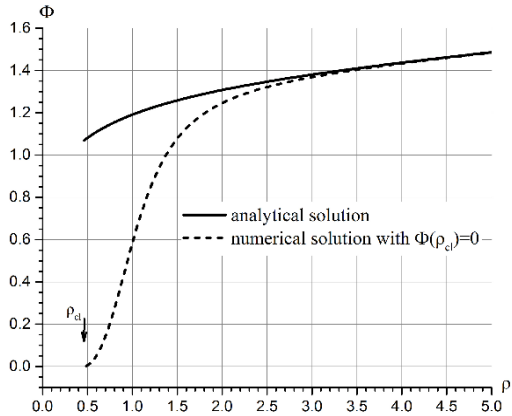
a) functions $\Phi(\rho)$

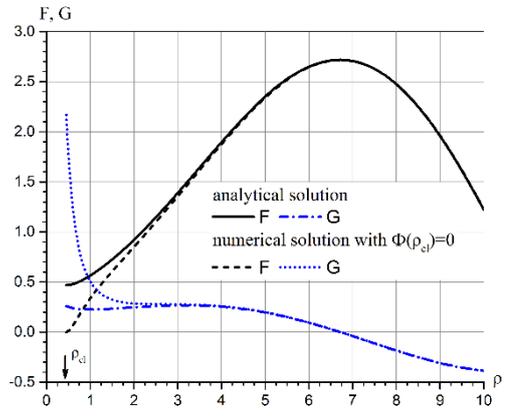
b) functions $F(\rho), G(\rho)$

Fig. 10. Dependences $\Phi(\rho), F(\rho), G(\rho)$ for the parameters
$\varepsilon = 1.2$, $\kappa = -1$, $Z = 137$, $\rho_{cl} = 0.45443$.

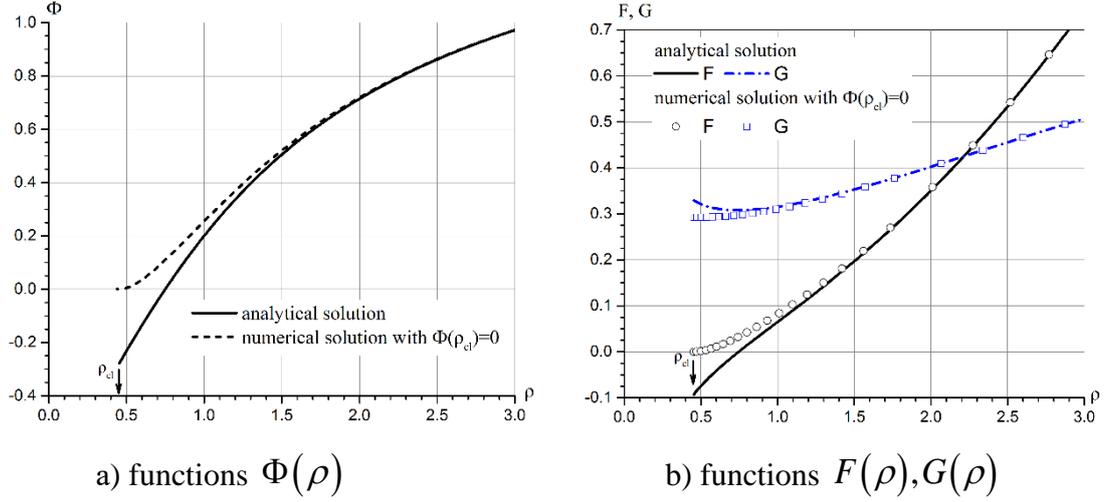

a) functions $\Phi(\rho)$

b) functions $F(\rho), G(\rho)$

Fig. 11. Dependences $\Phi(\rho), F(\rho), G(\rho)$ for the parameters
$\varepsilon = 1.2,\ \kappa = +1,\ Z = 137,\ \rho_{cl} = 0.45443$.

It is seen in the figures that at all considered values of the parameter there are differences between the dependences obtained with the boundary condition $\Phi(\rho_{cl}) = 0$ and the analytical dependences obtained using Eqs. (29).

However, these differences only exist at distances $\Delta$ of a few fractions of to several Compton wavelengths of the electron from $\rho_{cl}$. The largest differences are at $Z = 137$ and $\kappa = -1$ (see Fig. 10). The smallest differences are at $Z = 1$ and $\kappa = +1$ (see Fig. 7).

For all values of the considered parameters, the calculated dependences are very close to the analytical ones at $\rho \approx \rho_{cl} + \Delta$.

The above consideration shows that the use of new solutions of the Dirac equation will not lead to any new results as compared to the results obtained with the traditional Coulomb functions of the continuous spectrum (see (29)). In this sense, our consideration is only methodological. However, we have come to an important conclusion. In the repulsive Coulomb field, the domain of the stationary-state wave function of the Dirac equation and the relativistic Schrödinger-type equation does not include the region $0 \le \rho < \rho_{cl}$. In this connection, the use of the Schrödinger-type equation with the effective potential in quantum mechanics is somewhat analogous to introduction of the classical electron radius in classical physics. However, in quantum theory, an impermeable barrier occurs only for the fermion in the repulsive Coulomb field [11]. In the attractive Coulomb field, there is no such barrier. Neither is there an impermeable barrier at $\rho = \rho_{cl}$ for spinless particles. The half-integer spin and the certain sign of the electrical charge of fermions that lead to forming the quantum-mechanical impermeable barrier in the relativistic

Schrödinger-type equation should be explained by a future relativistic model of the fermion structure.

## 5. Conclusions

In quantum mechanics, motion of spin-½ particles in external electromagnetic and gravitational fields can be described using the Dirac equation with the bispinor wave function or the self-conjugate relativistic Schrödinger-type equation with the spinor wave function. The wave functions of both equations are connected with each other by relations (25), (26).

When the Schrödinger-type equation is used, in the effective potential of the repulsive Coulomb field there is a quantum-mechanical impermeable barrier with a radius $r_{cl} = \dfrac{Z(e^2/mc^2)}{1+E/mc^2}$. When the Dirac equation is used, there is no such barrier.

In the vicinity of $r_{cl}$, the Schrödinger-type equation has two solutions: $(F_{Schr})_1\big|_{r \to r_{cl}} \sim (r-r_{cl})^{3/2}, (F_{Schr})_2\big|_{r \to r_{cl}} \sim (r-r_{cl})^{-1/2}$. The second solution is divergent and nonrenormalizable at $r - r_{cl}$, and therefore it is physically unacceptable.

From the relations between functions (25) and (26), one can obtain two solutions of the Dirac equation in the vicinity of $r_{cl}$

$$\begin{cases} F_1\big|_{r\to r_{cl}} \sim (r-r_{cl})^2, \\ G_1\big|_{r\to r_{cl}} \sim \text{const 3}, \end{cases} \quad \begin{cases} F_2\big|_{r\to r_{cl}} \sim \text{const 1}, \\ G_2\big|_{r\to r_{cl}} \sim \text{const 2}. \end{cases}$$

To the solutions $F_2, G_2$ there correspond analytical Coulomb functions of the continuous spectrum. These functions are normally used in quantum-mechanical calculations. The functions corresponding to the solutions $F_1, G_1$ have not been investigated in the literature so far. Our analysis has shown that these solutions do exist. However, they diverge when $r \to 0$.

A paradoxical situation arises. To the physically acceptable solutions of the Dirac equation $F_2, G_2$ there corresponds the divergent solution of the Schrödinger-type equation $(F_{Schr})_2$. On the other hand, to the acceptable solution $(F_{Schr})_1$ there correspond the divergent solutions of the Dirac equation $F_1, G_1$.

We proposed the following way out. The impermeable barrier in the effective potential of the Schrödinger-type equation does not allow fermions to penetrate the region $r < r_{cl}$. In this case, the range $0 \leq r < r_{cl}$ is excluded from the wave function domain.

From the duality of the Dirac equation and the Schrödinger-type equation it follows that the range $0 \leq r < r_{cl}$ should also be excluded with necessity from the Dirac wave function domain. In this case, the problem of wave function divergence when $r \to 0$ disappears.

According to data available to us, a decrease in the wave functions domain does not lead to any new practical results. Earlier we showed that existence of the impermeable barrier did not affect the cross section of the Coulomb electron scattering in the lowest order of perturbation theory [11]. Owing to the dependence $r_{cl} \sim 1/E$ at $E \gg m$, the presence of the barrier does not contradict the results of the experiments on the study of the electron structure (see, for example, [11], [29]). The presented calculations with charges in the physical parameters show that new solutions of the Dirac equation with the boundary conditions $\Phi(r_{cl}) = 0$, $F(r_{cl}) = 0$ at distances of fractions or units of the Compton wavelength of the fermion from $r = r_{cl}$ almost coincide with the standard Coulomb functions of the continuous spectrum. Obviously, the use of the new solutions will not lead to any new results. On the other hand, our consideration is methodological and helpful for discussions on future development of quantum theory.

## Acknowledgements

The authors are grateful to A.L.Novoselova for her substantial technical assistance in preparation of the paper.